\documentclass[aps,prb,superscriptaddress,preprint]{revtex4}
\usepackage{graphicx} 
\usepackage{amsmath} 
\usepackage{multirow} 
\usepackage[breaklinks,colorlinks = true,linkcolor = blue,urlcolor=blue,citecolor=blue]{hyperref}

\begin{document}
\title{Peculiarities of optical absorption spectra of NdFe$_3$(BO$_3$)$_4$ crystal in magnetically ordered state and at the transition to spiral magnetic phase}

\author{V.V. Slavin}
\affiliation{B. Verkin Institute for Low Temperature Physics and 
Engineering of the National Academy of Sciences of Ukraine,
Nauky Ave., 47, Kharkiv, 61103, Ukraine}

\author{I.S. Kachur}
\affiliation{B. Verkin Institute for Low Temperature Physics and 
Engineering of the National Academy of Sciences of Ukraine,
Nauky Ave., 47, Kharkiv, 61103, Ukraine}

\author{V.S. Kurnosov}
\affiliation{B. Verkin Institute for Low Temperature Physics and 
Engineering of the National Academy of Sciences of Ukraine,
Nauky Ave., 47, Kharkiv, 61103, Ukraine}

\author{ V.G. Piryatinskaya}
\affiliation{B. Verkin Institute for Low Temperature Physics and 
Engineering of the National Academy of Sciences of Ukraine,
Nauky Ave., 47, Kharkiv, 61103, Ukraine}

\begin{abstract}
We study experimentally and theoretically optical absorption spectra of neodymium ferroborate in the temperature range $6-32$ K. We show that the temperature dependences of integral intensities of absorption lines demonstrate qualitatively different behavior for $\sigma$- and $\pi$-polarizations of the light. We suggest that this behavior may be caused by the appearance of a spiral magnetic structure in the compound under study. To describe the temperature dependences of lines intensities we propose a theoretical model which takes into account the Dzyaloshinskii-Moriya interaction between Nd$^{3+}$ and Fe$^{3+}$ ions.
\end{abstract}
\maketitle

\section{Introduction}
The study of rare-earth ferroborates with the general formula RFe$_3$(BO$_3$)$_4$ (R is a rare-earth element) is of interest due to the wide variety of magnetic structures and nontrivial phase transitions caused by the presence of two interacting magnetic subsystems, as well as due to the fact that many of the ferroborates exhibit multiferroic properties. 

All representatives of this family are antiferromagnets, in which magnetic ordering is determined by exchange interaction within the subsystem of Fe$^{3+}$ ions. The rare-earth subsystem acquires a magnetic moment due to exchange $f-d$ interaction with the iron ions. Rare earth ions usually have stronger magnetic anisotropy than Fe$^{3+}$ and determine the type of magnetic structure of the crystal. The crystal structure of the ferroborates is described at high temperatures by the space symmetry group $R32$; with decreasing temperature, a number of the crystals exhibit a structural phase transition to a lower symmetry phase $P3_121$ \cite{ref1}. 

Neodymium ferroborate NdFe$_3$(BO$_3$)$_4$ preserves the $R32$ crystal structure at least down to the temperature of 2 K \cite{ref2,ref3}. The rare-earth ion occupies a position with local symmetry $D_3$. At the temperature $T_N \approx 30$ K, antiferromagnetic ordering of the crystal occurs, and the magnetic moments of Fe$^{3+}$ and Nd$^{3+}$ ions are oriented along one of the second-order axes in the basal plane \cite{ref3,ref4}. Thus, three types of equivalent magnetic domains can arise. Below the temperature $T_{IC}\approx 13.5-16$ K, the collinear magnetic structure is transformed into a long-period antiferromagnetic spiral propagating along the $C_3$ axis \cite{ref5,ref6}, the orientation of the magnetic moments remaining parallel to the basal plane. At the same time, as shown in Ref. \cite{ref6}, the collinear phase does not disappear completely, but it coexists with the spiral phase down to the lowest temperatures. 

The electronic states of the Nd$^{3+}$ ion (configuration 4$f^3$) in the crystal field of $D_3$ symmetry are Kramers doublets, which are split at magnetic ordering of the crystal due to the exchange interaction with the iron ions subsystem. The value of exchange splitting of the ground Nd$^{3+}$ doublet is 8.8 cm$^{-1}$ at the temperature of 4.2 K \cite{ref7}. The energy difference between the ground and the first excited levels of the lowest multiplet $^4I_{9/2}$ is 65 cm$^{-1}$  \cite{ref8}. 

We have carried out for the first time the studies of optical absorption spectra of the NdFe$_3$(BO$_3$)$_4$ crystal in external magnetic fields of various directions with a field strength up to 65 kOe \cite{ref9,ref10}. In the geometry of transverse Zeeman effect, a nontrivial behavior of the intensities of polarized splitting components of the doublet line 15971-15978 cm$^{-1}$ in the region of the $^4I_{9/2}\to$ $^2H_{11/2}$ transition in Nd$^{3+}$ was discovered. We constructed a semi-empirical model that made it possible to describe the field dependences of the intensities of the Kramers doublet components \cite{ref9}. It was shown that the observed effect is a reflection of a complex pattern of effective fields that act on the Nd$^{3+}$ ion. These fields are a superposition of external (magnetic) and internal (exchange) fields, and the latter can have different values for the ground and optically excited states of Nd$^{3+}$ ion. In addition, it turned out that for an adequate description of the experiment it is necessary to assume that the Fe-Nd exchange interaction in this excited state has the opposite sign with respect to the exchange in the ground state. 

In order to confirm this assumption, we made a group-theoretical analysis of the electron transitions of Nd$^{3+}$ in the magnetically ordered state of the crystal, results of which were presented in brief report \cite{ref11}. Due to the fact that the easy-plane magnetic ordering leads to a decrease of local symmetry of the rare-earth ion position, new selection rules for transitions between the components of Kramers doublets arise. Analysis of the polarization properties of optical transitions makes it possible to determine the order of levels in the excited Kramers doublet of Nd$^{3+}$ and, correspondingly, the sign of Fe-Nd exchange interaction. In Ref. \cite{ref11} we showed that in the region of $^4I_{9/2}\to$ $^4G_{5/2}$  transition, there exist the neodymium excited states in which the Nd-Fe exchange interaction has the opposite sign with respect to the exchange in the ground state. In current work, we present the details of this group-theoretical analysis.

It is interesting to trace not only the transformation of spectrum at magnetic ordering of the crystal, but also its possible changes during the transition from the collinear antiferromagnetic phase to the long-period spiral phase. In this paper we carry out the studies of absorption spectra of NdFe$_3$(BO$_3$)$_4$ in the region of optical transition $^4I_{9/2}\to$ $^2H_{11/2}$ in the temperature range of $6-32$ K and analyze the temperature dependences of the main parameters of the absorption lines. We propose the theoretical model, describing the temperature dependence of absorption spectra. This model assumes the existence of Dzyaloshinskii-Moriya (DM) interaction between Nd and Fe ions which leads to formation of spiral magnetic structure. An influence of DM interaction on spiral structure formation was discussed in several papers (see, e.g., \cite{ref12,ref13,ref14,ref15}). In this work we demonstrate the impact of DM interaction on absorption spectra of neodymium ferroborate NdFe$_3$(BO$_3$)$_4$. 

\section{Experimental details}
For optical studies we used single crystals of NdFe$_3$(BO$_3$)$_4$ grown from the melt solution. The samples were made in the form of plane-parallel plates with a thickness of $0.2-0.7$ mm, oriented in such a way that the third- and second-order crystallographic axes lie in the plane of the sample. For absorption spectra studies, we used a diffraction spectrometer DFS-13. The light intensity was measured by linear photodiodes array, which was a part of the optical multichannel analyzer. The spectral resolution in the investigated region was about 0.5 cm$^{-1}$. The absorption spectra were recorded for light propagating perpendicular to the $C_3$ axis, with the direction of the light {\bf E}-vector being parallel ($\pi$-spectra) or perpendicular ($\sigma$-spectra) to the $C_3$ axis. For the temperature measurements we used a liquid-helium cooled cryostat with an internal volume filled by the gaseous helium where the sample was placed. 

\section{Polarization selection rules for electron transitions between Kramers states in local symmetries $D_3$ and $C_2$}

Polarization of electric dipole transition between the initial state $|b \rangle$ and the excited state $|a \rangle$ is determined by the matrix element of the dipole moment between these states $\langle a|{\bf p}|b \rangle$. The transition is allowed, if the direct product $\Gamma_b\times\Gamma_a^*$ contains an irreducible representation (IR) of the corresponding basis. For $D_3$ group, the list of IR and corresponding basis functions is given in the Table~\ref{table1}. IR $\Gamma_4$ and $\Gamma_{5,6}$ belong to the representations of the double group, taking into account transformations in the spin space. These representations are of particular interest in connection with the Kramers character of the Nd$^{3+}$ ion. All possible transitions between Kramers states within $D_3$ symmetry are listed in the Table~\ref{table2}. 

\begin{table}
\begin{tabular}{|c|c|} 
\hline
IR $D_3$        &   basis \\
\hline
$\Gamma_1\, (A_1)$  &   $m_z$\\
\hline
$\Gamma_2\, (A_2)$  &   $p_z$\\
\hline
$\Gamma_3\, (E)$  &   $m_x,m_y,p_x,p_y$\\
\hline
$\Gamma_4$  &   $\varphi_{+1/2},\varphi_{-1/2}$\\
\hline
$\Gamma_5$  &   $i\varphi_{+3/2}+\varphi_{-3/2}$\\
\hline
$\Gamma_6$  &   $\varphi_{+3/2}+i\varphi_{-3/2}$\\
\hline
\end{tabular}
\caption{IR of the group $D_3$ and the corresponding bases. $\Gamma_4$ is doubly degenerate IR, $\Gamma_{5,6}$ are one-dimensional complex conjugate IR. ${\bf p}$ is the electric-dipole moment, ${\bf m}$ is the magnetic-dipole moment.}
\label{table1}
\end{table}

\begin{table}
\begin{tabular}{|c|c|c|} 
\hline
$\Gamma_b \times \Gamma_a^*$        &   ${\bf p}$       & ${\bf m}$ \\
\hline
$\Gamma_5 \times \Gamma_6^*=\Gamma_6 \times \Gamma_5^*=\Gamma_2$ & $p_z$ & $-$\\
\hline
$\Gamma_6 \times \Gamma_6^*=\Gamma_5 \times \Gamma_5^*=\Gamma_1$ & $-$ & $m_z$\\
\hline
$\Gamma_4 \times \Gamma_4^*=\Gamma_1+\Gamma_2+\Gamma_3$ & $p_x,p_y,p_z$ & $m_x,m_y,m_z$\\
\hline
$\Gamma_4 \times \Gamma_5^*=\Gamma_4 \times \Gamma_6^*=\Gamma_3$ & $p_x,p_y$ & $m_x,m_y$\\
\hline
\end{tabular}
\caption{Polarizations of transitions between the Kramers states in the crystal field of $D_3$ symmetry for electric- (${\bf p}$) and magnetic-dipole (${\bf m}$) processes.}
\label{table2}
\end{table}

With easy-plane magnetic ordering of the crystal, the local symmetry of the rare-earth ion decreases. Let us consider this process for $C_2$ symmetry, which is the maximum for this type of ordering. One of the second-order axes in the $D_3$ symmetry, lying in the basal plane, is preserved and becomes the only axis in the $C_2$ group. The IR of the $C_2$ symmetry group and their correlation with the IR of $D_3$ group are given in the Table~\ref{table3}. The polarizations of transitions between the Kramers states in the monoclinic group are given in the Table~\ref{table4}. It should be noted here that the lower $C_2$ symmetry removes the strict prohibition on observing some transitions with certain polarizations, however the main features of the spectra continue to be inherited from the $D_3$ symmetry of the paramagnetic group of the crystal, as indicated in the Table~\ref{table3}. In the magnetically ordered state, the Kramers degeneracy of the energies of states $\Gamma_3$ ($\varphi _{+1/2}$) and $\Gamma_4$ ($\varphi _{-1/2}$) is also removed, and their splitting occurs. 

\begin{table}
\begin{tabular}{|c|c|c|} 
\hline
IR $C_2$        &   basis       & IR $D_3$\\
\hline
$\Gamma_1(A)$   & $m_x,p_x$     & $\Gamma_1, \Gamma_3 \, (m_x,p_x)$ \\
\hline
\multirow{2}{*}{$\Gamma_2(B)$ } & $m_y,p_y$ & $\Gamma_2, \Gamma_3 \, (m_y,p_y)$\\\cline{2-3}
                
                & $m_z,p_z$     &$\Gamma_2\, (p_z), \Gamma_3$\\
\hline
$\Gamma_3$      & $\varphi_{+1/2}$     & $\Gamma_4, \Gamma_5$ \\
\hline
$\Gamma_4$      & $\varphi_{-1/2}$     & $\Gamma_4, \Gamma_6$ \\
\hline
\end{tabular}
\caption{IR of the $C_2$ group and the corresponding bases in geometry, where the principal axis coincides with the $x$ direction for $D_3$ symmetry. $\Gamma_3$, $\Gamma_4$ are one-dimensional complex conjugate IR. The third column shows the correlation with IR of the $D_3$ group, in brackets are the basis functions inherited from $D_3$ to $C_2$ directly.}
\label{table3}
\end{table}

\begin{table}
\begin{tabular}{|c|c|c|} 
\hline
$\Gamma_b \times \Gamma_a^*$        &   ${\bf p}$       & ${\bf m}$ \\
\hline
$\Gamma_3 \times \Gamma_4^*=\Gamma_4 \times \Gamma_3^*=\Gamma_2$ & $p_y, p_z$ & $m_y,m_z$\\
\hline
$\Gamma_3 \times \Gamma_3^*=\Gamma_4 \times \Gamma_4^*=\Gamma_1$ & $p_x$ & $m_x$\\
\hline
\end{tabular}
\caption{Polarizations of transitions between Kramers states within the framework of $C_2$ symmetry for electric-dipole (${\bf p}$) and magnetic-dipole (${\bf m}$) processes.}
\label{table4}
\end{table}

As follows from the table, transitions between states of the same type ($\Gamma_3\to \Gamma_3$, $\Gamma_4\to \Gamma_4$) must have polarization strictly along the antiferromagnetism vector, that is, the selected second order axis of the crystal. Transitions between states of different types ($\Gamma_3\leftrightarrow \Gamma_4$) will be polarized in the plane perpendicular to the second order axis. Their predominant polarization within this plane will still be determined primarily by the symmetry of the transition in $D_3$ group. 

\section{Determination of Fe-Nd exchange interaction signs in excited states of Nd$^{3+}$ by analyzing the polarization properties of NdFe$_3$(BO$_3$)$_4$ absorption spectra}
\label{Sec4}

The electronic states of Nd$^{3+}$ ion in the crystal field of $D_3$ symmetry are Kramers doublets, which are transformed according to the two-valued IR $\Gamma_4$ and $\Gamma_{5,6}$. The ground multiplet $^4I_{9/2}$ is split into five Kramers doublets in the trigonal field: $^4I_{9/2} \to 3\Gamma_4 + 2\Gamma_{5,6}$, with the lowest level having $\Gamma_4$ symmetry \cite{ref8}. In accordance with the selection rules for $D_3$ symmetry (Table ~\ref{table2}), the electric-dipole transitions $\Gamma_4 \to \Gamma_4$ should be observed in both $\sigma$- and $\pi$-polarization, while the transitions $\Gamma_4 \to \Gamma_{5,6}$ - only in $\sigma$-polarization. 

\begin{figure}[ht]
\center{\includegraphics[width=10.0cm]{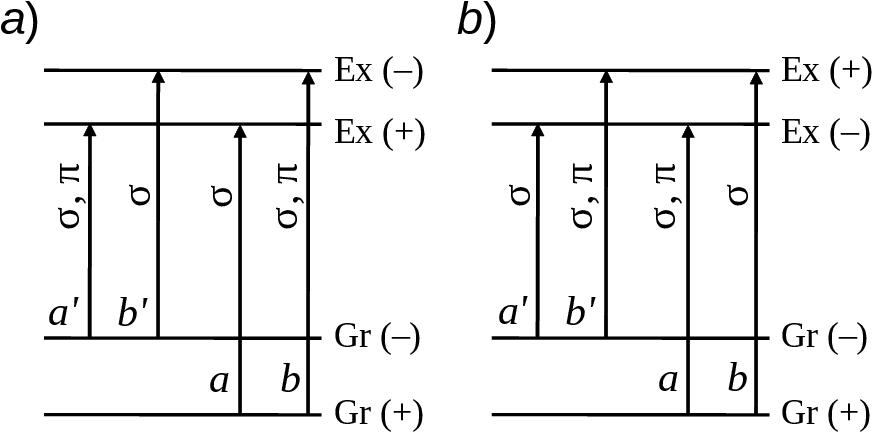} }
\caption{Schemes of transitions between the splitting components of the ground and excited Kramers doublets of Nd$^{3+}$ and polarization of electric-dipole transitions in $C_2$ symmetry for different arrangement of the excited doublet sublevels.}
\label{fig1}
\end{figure}

At temperatures $T<T_N$, the Kramers doublets of $\Gamma_4$ symmetry (including the ground doublet) are split due to the exchange interaction with the iron ions subsystem; as a result, up to four transitions can be observed, as shown in the schemes in Fig.~\ref{fig1}. At a temperature of 2 K, only the transitions $a$ and $b$ from the lowest sublevel of the ground doublet can be observed. With increasing temperature, the transitions $a'$ and $b'$ from upper sublevel of the ground doublet also appear. As for the states of $\Gamma_{5,6}$ symmetry, they do not split in the exchange field directed in the basal plane of the crystal, since they have $g_\perp \equiv 0$ for symmetry reasons. 

From the selection rules (Table \ref{table4}) it follows that at $T<T_N$, electric-dipole transitions between states of the same type can be observed only in $\sigma$-polarization, and transitions between states of different types can be observed in both $\sigma$- and $\pi$-polarization. Thus, if the order of the levels in the excited doublet is the same as in the ground one (Fig.~\ref{fig1}a), then the transitions $a'$ and $b$ will appear in $\pi$-polarization. If the order of the levels in the excited doublet is reversed, then in $\pi$-polarization the transitions $a$ and $b'$ will be observed (Fig.~\ref{fig1}b). The latter case corresponds to different signs of the exchange interaction with the iron subsystem in the ground and excited states of neodymium. 

\begin{figure}[ht]
\begin{minipage}[ht]{0.49\linewidth}
\center{\includegraphics[width=1.0\linewidth]{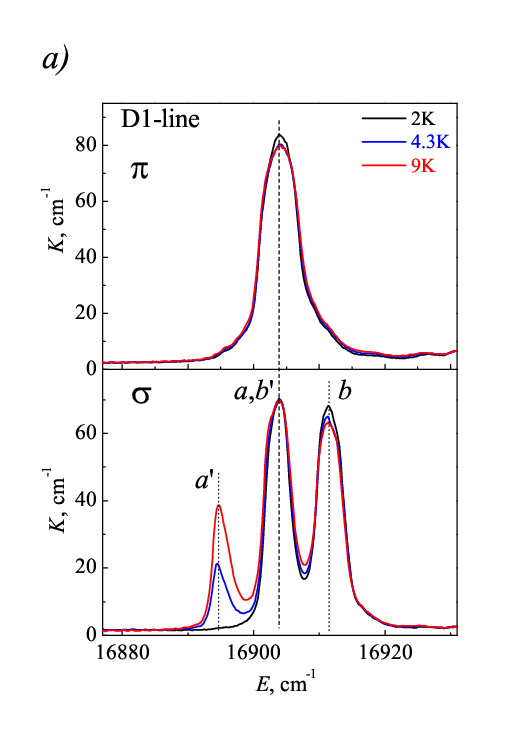} }
\end{minipage}
\begin{minipage}[ht]{0.49\linewidth}
\center{\includegraphics[width=1.0\linewidth]{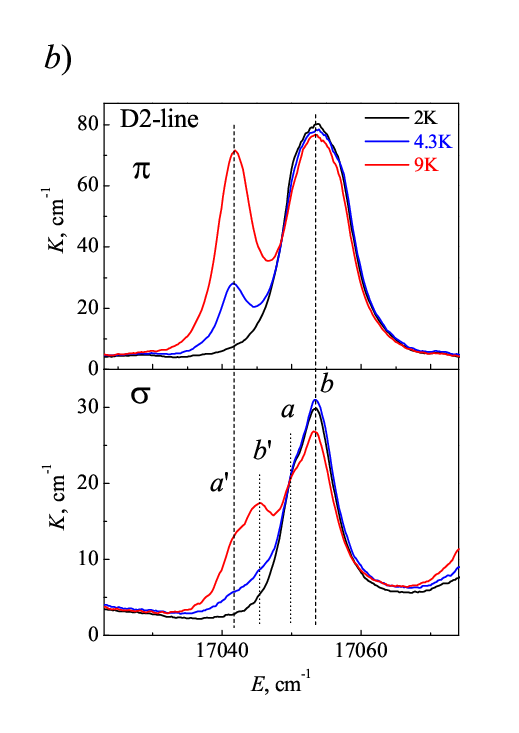} }
\end{minipage}
\caption{Low-temperature absorption spectra of NdFe$_3$(BO$_3$)$_4$ in the region of lines D1 (a) and D2 (b) of the optical transition $^4I_{9/2}\to$ $^4G_{5/2}$ (D-group) in $\pi$- and $\sigma$- polarizations \cite{ref11}.}
\label{fig2}
\end{figure}

The polarization properties of optical absorption spectra confirm the existence of excited states of Nd$^{3+}$ with an inversion of the sign of the Fe-Nd exchange interaction. So, Fig.~\ref{fig2} presents the low temperature spectra of NdFe$_3$(BO$_3$)$_4$ crystal in the region of the D1 and D2 lines of the optical transition $^4I_{9/2}\to$ $^4G_{5/2}$ (D-group) in $\sigma$- and $\pi$-polarizations \cite{ref11}. The spectrum of D-group in $\alpha$-polarization completely coincides with the $\sigma$-polarized spectrum, which testifies to the electric-dipole nature of the optical transitions. The lines D1 and D2 correspond to transitions of the $\Gamma_4 \to \Gamma_4$ type in $D_3$ symmetry. 

In low temperature spectra of the D1 line, three exchange splitting components can be observed in $\sigma$-polarization (Fig.~\ref{fig2}a). The value of exchange splitting of the excited state, equal to the energy difference between the transitions $a$ and $b$, is approximately 7.7 cm$^{-1}$ at $T$ = 2 K, which is quite close to the splitting value of the ground doublet - 8.8 cm$^{-1}$ \cite{ref7}. That is why the transitions $a$ and $b'$ have close energies and are observed in the spectra as a single line. Just these transitions manifest themselves in $\pi$-polarization, which corresponds to the diagram in Fig.~\ref{fig1}b. Therefore, the Fe-Nd exchange interaction for this excited state of Nd$^{3+}$ has the opposite sign with respect to the exchange in the ground state. 

For the D2 line (Fig.~\ref{fig2}b), splitting value of the excited doublet is approximately 3.7 cm$^{-1}$ at 2 K. With increasing temperature, all four possible transitions can be observed in the $\sigma$-spectrum. But the picture observed in $\pi$-polarization is essentially different from that described above: in this case, the lowest-frequency transition $a'$ and the highest-frequency one $b$ are manifested. Such polarization of transitions is in accordance with the scheme in Fig.~\ref{fig1}a, in which the order of sublevels (and the sign of exchange interaction) in the ground and excited doublets are the same. 

\section{Temperature evolution of the absorption spectra of NdFe$_3$(BO$_3$)$_4$ in the region of magnetic ordering of the crystal}

We have traced the evolution of NdFe$_3$(BO$_3$)$_4$ absorption spectra in the temperature range corresponding to the magnetically ordered state of the crystal in order to identify its possible changes during the transition from the collinear antiferromagnetic phase to the spiral one. The temperature studies of the spectra were carried out in the region of the optical transition $^4I_{9/2}\to$ $^2H_{11/2}$ (C-group) in Nd$^{3+}$. This spectral region was chosen due to the relatively small (compared to D-group) intensity of absorption, which is important for correct determination of the main parameters of the spectral lines.

The $^2H_{11/2}$ multiplet is split in the crystal field of $D_3$ symmetry into six Kramers doublets: $^2H_{11/2}\to 4\Gamma_4 + 2\Gamma_{5,6}$. In Ref. \cite{ref10} we found and identified all six transitions from the ground state to the levels of the $^2H_{11/2}$ multiplet in the absorption spectrum of NdFe$_3$(BO$_3$)$_4$. The values of splitting of neodymium excited states in the exchange field, as well as the values of $g$-factors of these states were determined. The energies of the optical transitions at a temperature 35 K were as follows: 15805 cm$^{-1}$ (C1), 15837 cm$^{-1}$ (C2), 15842 cm$^{-1}$ (C3), 15932 cm$^{-1}$ (C4), 15971 cm$^{-1}$ (C5) and 15997 cm$^{-1}$ (C6) \cite{ref10}. 

\begin{figure}[ht]
\begin{minipage}[ht]{0.49\linewidth}
\center{\includegraphics[width=1.0\linewidth]{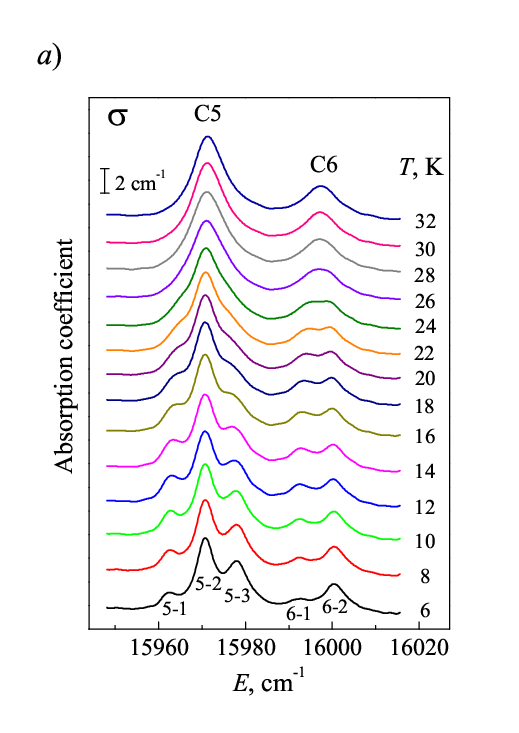} }
\end{minipage}
\begin{minipage}[ht]{0.49\linewidth}
\center{\includegraphics[width=1.0\linewidth]{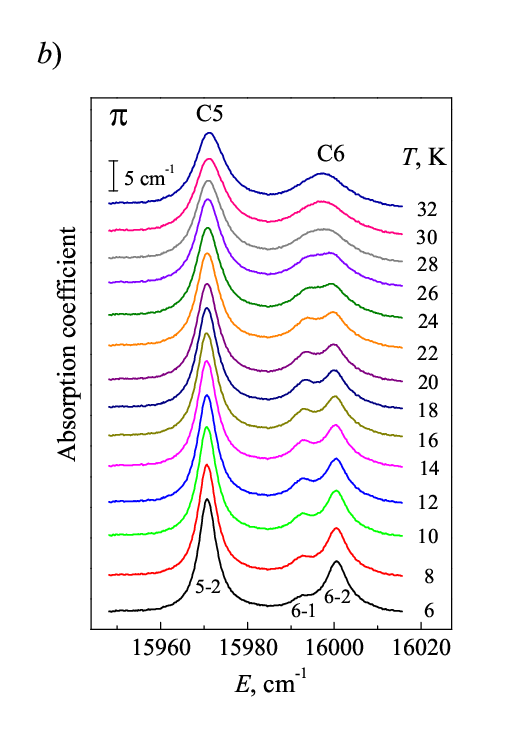} }
\end{minipage}
\caption{Absorption spectra of NdFe$_3$(BO$_3$)$_4$ in the region of C5 and C6 lines of the optical transition $^4I_{9/2}\to$ $^2H_{11/2}$ (C-group) in $\sigma$(a) and $\pi$(b) polarizations at different temperatures. }
\label{fig3}
\end{figure}

Figure \ref{fig3} shows the temperature series of the optical absorption spectra of NdFe$_3$(BO$_3$)$_4$ in the region of C5 and C6 lines in $\sigma$- and $\pi$-polarizations. In the C-group, as well as in D-group, all transitions are of electric-dipole character \cite{ref10}. Both lines C5 and C6 are associated with transitions of the type $\Gamma_4 \to \Gamma_4$ in the crystal field $D_3$. 

For the C5 line, the exchange splitting of the excited state at T = 2 K is approximately 7.5 cm$^{-1}$, which is quite close to the splitting value of the ground doublet. Therefore, similarly to the case of the D1 line, at temperatures $T<T_N$ three splitting components are observed in the $\sigma$-spectra of the C5 line. In $\pi$-polarization, the central transition $a,b'$ appears (line 5-2), which indicates the reverse order of levels arrangement in the excited doublet with respect to the ground one (Fig.~\ref{fig1}b), and, consequently, the inversion of the Fe-Nd interaction sign. (Note that the same result was obtained by us in Ref. \cite{ref9} on the base of theoretical description of the field dependences of the intensities of this doublet components in transverse Zeeman geometry). In the case of the C6 line, the excited state shows practically no splitting in the exchange field \cite{ref10}. The two splitting components 6-1 and 6-2 observed in the $\sigma$- and $\pi$-spectra of this line are associated with the transitions from the upper and lower sublevels of the ground doublet to the unsplit excited level ($a'$+$b'$ and $a$+$b$). 

It is seen from Fig.~\ref{fig3} that no significant changes in the shape of the spectra are observed in the temperature range of 13.5-16 K, corresponding to the phase transition from collinear easy-plane magnetic structure to the spiral one. Therefore, we traced the temperature evolution of the main parameters of the absorption lines, namely the position of the maxima, halfwidth and integral intensity. For this purpose, the absorption spectra were decomposed into Lorentzian components. 

\begin{figure}[ht]
\center{\includegraphics[width=10.0cm]{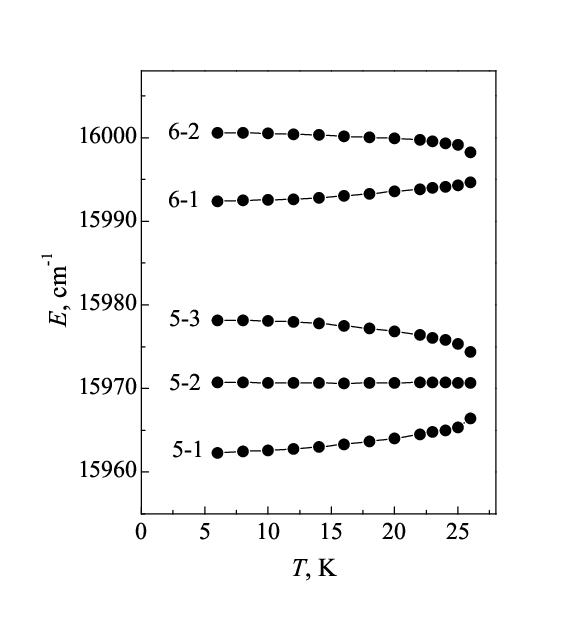} }
\caption{Energies of the splitting components of C5 and C6 lines of the optical transition $^4I_{9/2}\to$ $^2H_{11/2}$ depending on temperature.}
\label{fig4}
\end{figure}

Figure \ref{fig4} presents the temperature dependences of energies of the maxima of splitting components of C5 and C6 lines. It is seen that these dependences are monotonic and have no peculiarities in the region of $T_{IC}$. This can be explained by the fact that, due to rather weak magnetic anisotropy in the basal plane, the energy characteristics of optical transitions change very little when the magnetic moments are rotated in this plane. As the temperature approaches $T_N$, the converging of the components of the exchange splitting occurs. The temperature dependences of the line components half-widths also exhibit a monotonic character. 

\begin{figure}[ht]
\begin{minipage}[ht]{0.49\linewidth}
\center{\includegraphics[width=1.0\linewidth]{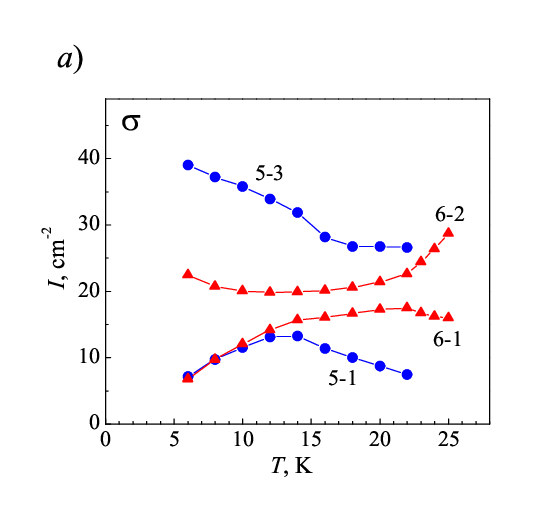} }
\end{minipage}
\begin{minipage}[ht]{0.49\linewidth}
\center{\includegraphics[width=1.0\linewidth]{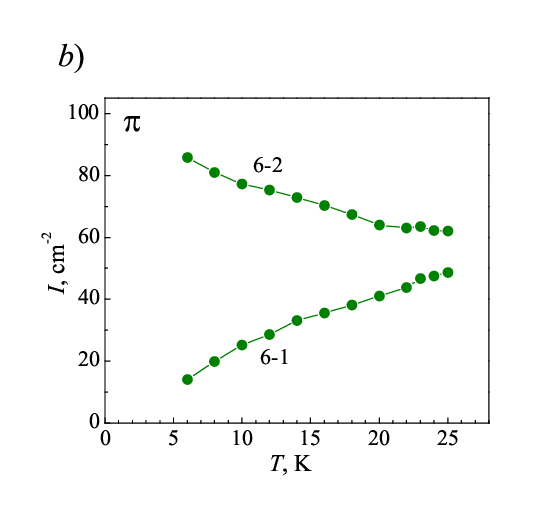} }
\end{minipage}
\caption{Integral intensities of the splitting components of the lines C5 and C6 in $\sigma$-polarization (a) and of the line C6 in $\pi$-polarization (b) depending on temperature. }
\label{fig5}
\end{figure}

Figure \ref{fig5} shows the temperature dependences of the integral intensities of the splitting components of C5 and C6 lines. Such dependences for “cold” and “hot” transitions should be determined by the temperature dependences of populations of the lower and upper sublevels of the ground neodymium doublet, from which, respectively, these transitions occur. In our case, lines 5-3 and 6-2 are transitions from the lower sublevel, lines 5-1 and 6-1 are transitions from the upper one, and line 5-2 is the sum of “cold” and “hot” components. However, the intensities of some components in $\sigma$-polarized spectra exhibit anomalous temperature dependences. So, the “hot” component 5-1 at $T > 14$ K exhibits a decrease in intensity with increasing temperature (Fig.~\ref{fig5}a), which cannot be explained by taking into account only the temperature dependence of the population. Both components of line C6 also reveal unusual behavior in $\sigma$-polarization in approximately the same temperature range (Fig.~\ref{fig5}a). These anomalies can be associated with a change of the directions of neodymium magnetic moments in the basal plane during the transition from the spiral structure to the collinear one. As for the behavior of the intensities in $\pi$-spectra, it does not reveal any obvious anomalies (Fig.~\ref{fig5}b), which is quite understandable, since rotation of the magnetic moments in the basal plane does not lead to a change in the probability of electric-dipole transitions in $\pi$-polarization. 

\begin{figure}[ht]
\center{\includegraphics[width=10.0cm]{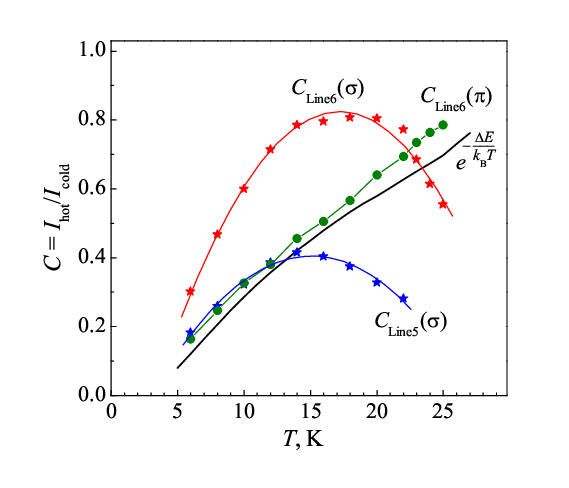} }
\caption{The ratio of integral intensities of the "hot" and "cold" components of C5 and C6 lines of the optical transition $^4I_{9/2} \to$ $^2H_{11/2}$ depending on the temperature. The black solid line corresponds to $\exp\left(-\frac{\Delta E}{k_BT}\right)$.  Solid red and blue curves are the results of fitting the experimental data by the expression \eqref{eq:C}. Solid green curve is a guide for eye.}
\label{fig6}
\end{figure}

It is interesting to depict the temperature dependence of such a quantity as the ratio of the intensities of “hot” and “cold” transitions which have the same symmetry. For example, components 5-1 and 5-3 of the line C5 correspond to the transitions $a'$ and $b$, both occurring between states of the same type (see diagram in Fig.~\ref{fig1}b). For line C6, components 6-1 and 6-2 also obey the same selection rules, since they are associated with transitions of types $a'+b'$ and $a+b$. In such cases, the ratio of the intensities of the “hot” and “cold” transitions should be determined only by the ratio of populations of the upper and lower sublevels of the ground doublet. However, from Fig.~\ref{fig6} it is evident that only for $\pi$-polarized components of line C6 the relation $C(T)= I_{hot}/I_{cold} \sim exp(-\frac{\Delta E}{k_BT})$ (where $\Delta E$ is the splitting of the ground doublet, which temperature dependence was taken from Ref. \cite{ref7}) is approximately fulfilled. As for $\sigma$-polarization, for both line C5 and line C6, the $C(T)$ dependences change their character just in the temperature range that corresponds to the reorientation of magnetic moments from the spiral to the collinear structure.

\section{Theoretical Model}
To describe the temperature behavior of lines intensity in absorption spectra of NdFe$_3$(BO$_3$)$_4$ we propose the following model. We take into account exactly the interaction of Nd ion with crystal electric field (CEF), and the interaction between Nd and Fe ions. Fe-Fe ion interactions are considered in the framework of mean-field approximation. Hence, the Hamiltonian of the system under consideration has the form:
\begin{equation}
{\hat H}={\hat H}_{Nd} +{\hat H}_{Fe}+{\hat H}_{Fe-Nd}.
\label{eq1}
\end{equation}
Here ${\hat H}_{Nd}$ is the Hamiltonian describing the interaction of Nd ion with CEF: 
$${\hat H}_{Nd}=\alpha a_{20}B^0_2{\hat O}^0_2+\beta\left(a_{40}B_4^0{\hat O}_4^0+ a_{43}B_4^3{\hat O}_4^{-3}\right)+
\gamma\left(a_{60} B^0_6{\hat O}^0_6+a_{63} B^3_6{\hat O}^{-3}_6+a_{66} B^6_6{\hat O}^6_6\right).$$
Here ${\hat O}_i^j$ are the Stevens operators \cite{ref16}, $B_i^j$ are the CEF parameters. According to Ref. \cite{ref8} for NdFe$_3$(BO$_3$)$_4$ we have: 
$$B^0_2=551, \, B^0_4=-1239, \, B^3_4=697, \, B^0_6=519, \, B^3_6=105, \, B^6_6=339.$$
The constants $a_{ij}$ are:
$$a_{20}=1/2, \, a_{40}=1/8, \, a_{43}=\sqrt{35}/2, \, a_{60}=1/16, \, a_{63}=\sqrt{105}/8, \, a_{66}=\sqrt{231}/16$$
and the parameters $\alpha,\beta,\gamma$ for the ground multiplet $^4I_{9/2}$ and the excited multiplet $^2H_{11/2}$ are presented in Table \ref{table5} (see Ref. \cite{ref16}).

\begin{table}
\begin{tabular}{|c|c|c|} 
\hline
& The ground  & The excited \\
&  multiplet $^4I_{9/2}$  & multipet $^2H_{11/2}$\\
\hline
$\alpha$    & $-\frac{7}{3^2\cdot 5}$ & $\frac{4\cdot \sqrt{14}}{11^2\cdot 13}$\\
$\beta$   & $-\frac{2^3\cdot 17}{3^3\cdot 11^3 \cdot 13}$ & $\frac{2^5\cdot 17\cdot \sqrt{14}}{3^3\cdot 7\cdot 11^3 \cdot 13}$\\
$\gamma$    & $-\frac{5\cdot 17\cdot 19}{3^3\cdot 7 \cdot 11^3 \cdot 13^2}$ & $\frac{2\cdot 5\cdot 19 \cdot \sqrt{14}}{3\cdot 7\cdot 11^3 \cdot 13^2}$ \\ 
\hline
\end{tabular}
\caption{The parameters $\alpha,\beta,\gamma$  for the ground and excited multiplets.}
\label{table5}
\end{table}

Instead of many-body Hamiltonian describing Fe-Fe ion interactions, we use mean-field model. In this model we take into account $N_{nb}=6$ Fe ions, surrounding the Nd ion. These Fe ions are in equivalent crystal position. We take into account the interactions of these Fe ions with the rest in the framework of mean-field model:
\begin{equation}
{\hat H}_{Fe}=-{\hat {\bf J}}_{Fe} h_{m-f}.
\label{eq:h_fe}
\end{equation}
Here  $\hat{\bf J}_{Fe}$ is the Fe ion moment (in dimensionless units), $h_{m-f}$ is the effective magnetic field, produced by the rest of Fe ions (in energy units).  The dependence of $h_{m-f}$ on temperature $T$ in the range of magnetic ordering of Fe subsystem has the form: 
\begin{equation}
 h_{m-f}(T)=h_0\tanh\left(\frac{T_c-T}{T_0}\right).
\label{eq:h_mf}
\end{equation}
In the proposed model we {\it explicitly} take into account the interactions of $N_{nb}$ ions of Fe with the rest of Fe subsystem. It is obvious that the Fe subsystem affects on Nd ions also. This interaction is responsible for the splitting of the Nd ground doublet with gap $\Delta E(T)$ and, hence, $\Delta E(T)$ is proportional to $h_{m-f}(T)$. That is why we approximated \eqref{eq:h_mf} by the experimental data for $\Delta E(T)$ \cite{ref7} and obtained $T_c$ and $T_0$. Parameter $h_0$ was fitted so that the splitting of Nd ground doublet $\Delta E(T)$ at $T=4.2$ K coincided with the experimentally measured value \cite{ref7}. 

Further, we consider the interaction between Fe subsystem and Nd ion {\it implicitly}, through the Hamiltonian ${\hat H}_{Fe-Nd}$ which describes the interaction between the Nd ion and  $N_{nb}$ ions of Fe in the framework of $xxz$ model:
\begin{equation}
 {\hat H}_{Fe-Nd}=N_{nb}\left\{J_{xx}\left(
 \hat {\bf J}_{Fe}^x\hat {\bf J}_{Nd}^x+
 \hat {\bf J}_{Fe}^y\hat {\bf J}_{Nd}^y
 \right)+J_{zz}\hat {\bf J}_{Fe}^z\hat {\bf J}_{Nd}^z+
 J_{DM}\left(\hat {\bf J}_{Fe}^x\hat {\bf J}_{Nd}^y-
 \hat {\bf J}_{Fe}^y\hat {\bf J}_{Nd}^z\right)\right\}.
 \label{eq:xxz}
\end{equation}

Here $\hat {\bf J}_{Nd}$ is Nd ion moment (in dimensionless units), $J_{xx}$, $J_{zz}$, and $J_{DM}$ are the exchange constants (in energy units). The first two terms describe exchange in $ab$ plane and $c$ direction, the last term describes Dzyaloshinskii-Moriya interaction (note, the role of DM interaction in spiral phase formation in neodymium ferroborate was discussed in \cite{ref5,ref15}).

The spectral data $\{\lambda_k,\phi_k\}$ of the Hamiltonian \eqref{eq1} corresponding to the ground multiplet $^4I_{9/2}$ were obtained numerically (here $\lambda_k, k=1,2, \ldots$ are the eigenvalues and $\phi_k$ are the eigenvectors). Spectral data  $\{\lambda '_k,\phi '_k\}$ corresponding to the excited multiplet $^2H_{11/2}$ were obtained similarly. Using the eigenvectors of the ground doublet $\phi_1$, $\phi_2$ we calculated the density matrices $\rho_1, \rho_2$. Note, due to the fact that the Hamiltonian  ${\hat H}_{Fe}$ depends on the temperature $T$ (see \eqref{eq:h_fe} and \eqref{eq:h_mf}), these density matrices depend on $T$ also: $\rho_1=\rho_1(T), \rho_2=\rho_2(T)$. Performing tracing over Fe ion variables we obtained the reduced density matrices of Nd ion: 

$$(\rho_{1,2})_{Nd}={\rm Sp}_{Fe}(\rho_{1,2}).$$
The observable orientation of Nd moment in $ab$ plane is: 
$${\rm Sp}\left((\rho_{1,2})_{Nd}O^2_2\right)=\langle J_x^2-J_y^2 \rangle=J^2\cos(2\varphi_{1,2})$$
Here  $\varphi$ is the angle between mean value of ${\bf J}$  projection on $ab$ plane and $a$ axis. 

Similarly to $\rho_1,\rho_2$, for each line C5 and C6 we calculated the density matrices $\rho'_1, \rho'_2$ corresponding to the components of the excited doublet.  As the result we obtain: 
$${\rm Sp}\left((\rho'_{1,2})_{Nd}O^2_2\right)=\langle J_x^2-J_y^2 \rangle=J^2\cos(2\varphi'_{1,2}).$$

The intensities of lines $I_{\rm{hot}}$, $I_{\rm{cold}}$ are proportional to the population of the corresponding states and to the square of the modulus of dipole transition matrix element. Then for the ratio  
$C=\frac{I_{\rm{hot}}}{I_{\rm{cold}}}$ we obtain: 
\begin{equation}
C=\frac{I_{\rm{hot}}}{I_{\rm{cold}}}\sim\left(\frac{\cos(\varphi_2-\varphi'_1)}{\cos(\varphi_1-\varphi'_2)}\right)^2 \exp\left(-\frac{\Delta E}{k_B T}\right).
\label{eq:C}    
\end{equation}

Here $k_B$ is the Boltzmann constant. Fitting the experimental dependencies of $C_{\rm{Line}\, 5,6}(T)$ gives the following values of exchange constants (see Table \ref{table6}). In this model we obtain information only about the absolute values of the exchange constants of excited states. Indeed, a change in the signs of the exchange constants is equivalent to a change in the directions of the Nd ion moments by 180$^0$, i.e. the order of the sublevels in the doublets (see Section \ref{Sec4}).

\begin{table}
\begin{tabular}{|c|c|c|c|} 
\hline
Exchange constants  & The ground  & The excited multiplet
& The excited multiplet
\\
(cm$^{-1}$) & multiplet $^4I_{9/2}$   & 
$^2H_{11/2}$, line C5  & $^2H_{11/2}$, line C6 \\
\hline
$|J_{xx}|$    &   0.36    &  0.22 & 0.08\\
$|J_{zz}|$    &   0.34    &  0.16 & 0.05\\
$|J_{DM}|$    &  0.023    &  0.01 & 0.009\\
\hline
\end{tabular}
\caption{Optimal values of exchange constants in the Hamiltonian \eqref{eq:xxz}.}
\label{table6}
\end{table}

The results of fitting are presented in Figure \ref{fig6}. As can be seen, the proposed theoretical model describes the experimental data well. It should be noted that {\it without} taking into account the DM interaction, it is impossible to describe the experimental dependencies even {\it qualitatively}. Thus, we have shown that it is DM interaction that is responsible for the unusual temperature dependence of the light absorption coefficient.

\section{Summary}

We performed complex experimental and theoretical studies of the optical absorption spectra of neodymium ferroborate NdFe$_3$(BO$_3$)$_4$ in the region of the transition $^4I_{9/2}\to$ $^2H_{11/2}$ in the temperature range $6-32$ K for $\sigma$ (${\bf k}||C_3, {\bf E}\perp C_3$) and $\pi$ (${\bf k}||C_3, {\bf E}||C_3$) polarizations of light. We made a group-theoretical analysis of electron transitions of Nd$^{3+}$ in the magnetically ordered easy-plane state of the crystal which made it possible to determine the signs of Fe-Nd exchange interaction in the excited states of neodymium. We obtained the temperature dependences of the main parameters of spectral lines for both $\pi$- and $\sigma$- light polarizations. We have shown that for $\pi$-polarization the temperature dependences of integral intensity of absorption lines can be described well in terms of thermoactivation model. In this model the integral intensities of lines are determined by the thermal populations of the lowest and excited sublevels of the ground doublet of $^4I_{9/2}$ multiplet. At the same time the temperature behavior of lines intensity in $\sigma$-spectra demonstrates crucial deviation from the thermoactivation model. We explain this distinction in temperature dependences by the phase transition and spiral phase appearance. Since $z$-components of Nd$^{3+}$ ion moments are absent in the whole temperature range $T<T_N$, $\pi$-spectra are non-sensitive to this phase transition. In $\sigma$-spectra the appearance of non-collinear (spiral) phase manifests itself in full extent. Due to the rotation of Nd$^{3+}$ ion moment in $ab$ plane the probability of optical transitions depends both on temperature population of the corresponding energy levels and on the rotation angle of this moment. We proposed theoretical model that takes into consideration the DM interaction between Nd$^{3+}$ and Fe$^{3+}$ ions, which causes rotation of magnetic moments. Our calculations showed good agreement between the experimental and theoretical data. It confirmed the key role of DM interaction in the formation of such unusual temperature dependences of integral intensities of absorption lines. Moreover, without DM interaction, it is impossible to describe the experimental data even qualitatively.

\end{document}